\documentclass[conference]{IEEEtran}
\IEEEoverridecommandlockouts
\usepackage{cite}
\usepackage{amsmath,amssymb,amsfonts}
\usepackage{graphicx}
\usepackage{textcomp}
\usepackage{xcolor}
\usepackage{url}
\usepackage{booktabs}
\usepackage{array}
\usepackage{algorithm}
\usepackage{algpseudocode}

\def\BibTeX{{\rm B\kern-.05em{\sc i\kern-.025em b}\kern-.08em
    T\kern-.1667em\lower.7ex\hbox{E}\kern-.125emX}}

\begin{document}

\title{Optimizing API Gateway Placement in Multi-Cloud Kubernetes}

\author{
\IEEEauthorblockN{ Vinoth Punniyamoorthy}
\IEEEauthorblockA{\textit{Texas, USA} \\
 0009-0009-3719-4949}
 \and
 \IEEEauthorblockN{Murali Shankar Dulam}
\IEEEauthorblockA{\textit{Texas, USA} \\
 0009-0000-7231-834X}
 \and
\IEEEauthorblockN{ Aswathnarayan Muthukrishnan Kirubakaran}
\IEEEauthorblockA{\textit{California, USA} \\
 0009-0006-6652-2663}
 \and
\IEEEauthorblockN{Akshay Deshpande}
\IEEEauthorblockA{\textit{California, USA } \\
0009-0002-3007-3393}
\and

\IEEEauthorblockN{Nachiappan Chockalingam	}
\IEEEauthorblockA{\textit{Massachusetts, USA} \\
 0009-0007-4275-3771}
\and
\IEEEauthorblockN{Bikesh Kumar}
\IEEEauthorblockA{\textit{Texas, USA} \\
 0009-0009-7190-1862}
\and
\IEEEauthorblockN{Naga Surya Pasupuleti}
\IEEEauthorblockA{\textit{Texas, USA} \\
0009-0001-0823-2186}
\and
\IEEEauthorblockN{Narender Reddy	Bitla}
\IEEEauthorblockA{\textit{Texas, USA} \\
0009-0002-6862-3557}
}

\maketitle

\begin{abstract}
The use of API gateways within geographically distributed multi-cloud Kubernetes clusters poses a tradeoff between infrastructure cost, computational resources, and network latencies. We present an optimization formulation that addresses API gateway placement as a capacitated facility location problem that jointly determines which candidate clusters to activate, how many gateway replicas to deploy, and how regional traffic should be distributed across the selected clusters. The formulation imposes an upper bound on estimated client-to-cluster network round-trip latency, excluding gateway processing, queuing, and backend-service latency, and incorporates a utilization headroom factor for gateway replica capacity. We present both a mixed-integer linear programming (MILP) formulation and a constructive greedy heuristic that ranks candidates according to incremental cost, comprising cluster-activation and marginal replica costs, per unit of assignable capacity while explicitly accounting for already-committed load. Both formulations are applied to deterministic, seed-controlled, geography-based synthetic instances. For each problem size, 30 instances are generated with random seeds to analyze their performance. The greedy algorithm achieves an optimality gap of 3.2\% to 4.7\% to the MILP optimal solution, with a maximum observed gap of 25.0\% for one particular instance, and a speedup of approximately 660× to 3,490× for 3 to 12 candidate clusters. In a canonical 10-candidate, 10-demand region instance, MILP-optimal deployment saves 24.2\% in terms of monthly cost compared to the full-replication baseline. On the other hand, selecting the single cheapest candidate yields savings of 24.8\% compared to the MILP optimum but does not satisfy the latency requirement for 3 out of 10 demand regions. Cutting the latency budget in half leads to an increase in the modeled cost by 15.1\%, and the optimal number of clusters is expanded to 5 from 3. The results above are generated using synthetic planning instances, and hence they represent a proof-of-concept rather than indicative production cost or latency improvements.
\end{abstract}


\section{Introduction}
\label{sec:intro}

API gateways provide a common entry point for authentication, authorization, rate limiting, routing, observability, and policy enforcement in cloud-native systems. As organizations deploy applications across multiple Kubernetes clusters, cloud providers, and geographic regions, gateway replicas may need to be placed near distributed client populations to satisfy network-latency requirements. Activating gateways in additional clusters, however, introduces fixed operational expense and per-replica compute cost. Platform teams must therefore balance geographic proximity, available capacity, and infrastructure cost when planning gateway deployments.

This paper investigates the following static planning problem: given a set of candidate clusters, regional request demand, deployment costs, replica capacity, and estimated client-to-cluster network latency, which clusters should host gateway replicas, how many replicas should each selected cluster run, and how should regional traffic be assigned? The objective is to minimize monthly deployment cost subject to capacity and per-region latency constraints.

In this paper, $\ell_{ij}$ denotes an estimated client-to-cluster network round-trip time, not complete end-to-end application response time; gateway processing, queueing, load-balancer, TLS handshake, and backend-service latency are outside the model. The formulation is a static planning model, not a runtime routing or autoscaling mechanism: autoscaling adjusts replica counts within an already-selected cluster in response to live load, whereas this model decides which clusters should host capacity in the first place, on a planning cadence (e.g., quarterly). The MILP contains no redundancy, failure-domain, or cross-provider availability constraint, so no high-availability property should be inferred from a computed placement.

Existing studies have examined API gateway performance and cost
trade-offs on fixed deployment platforms
policy integration and enforcement in containerized
environments~\cite{kim2024kubeaegis}, and SLO-aware or
cost-sensitive autoscaling of containerized workloads within
individual Kubernetes clusters~\cite{marchese2025slo,vu2022predictive}.
However, these approaches do not jointly determine gateway placement
and replica allocation across heterogeneous multi-cloud Kubernetes clusters while explicitly considering both latency-SLA feasibility and deployment cost. None of these jointly optimize gateway location, replica count, and regional traffic assignment across candidate clusters. Capacitated facility location with distance- or latency-restricted assignment is a long-studied combinatorial problem \cite{cornuejols1990uflp} that underlies content-delivery replica placement \cite{sahoo2017replica,salahuddin2018contentplacement}, geographic $k$-center placement \cite{hillmann2020kcenter}, response-time-optimized distributed cloud resource allocation, geo-distributed data placement, virtual network function placement in multi-cloud service chains, and SLA- or QoS-aware service placement in edge-cloud continuum systems. We adapt this established problem structure to API gateway replica placement specifically; the contribution is this adaptation and its empirical evaluation, not a new class of optimization problem. Kubernetes-native scheduling surveys address a related but distinct problem: placing pods onto nodes within a cluster, rather than deciding which geographically distributed clusters should host a service at all.

This paper makes three significant contributions. First, we transform the multi-cloud API gateway placement into a capacitated facility location problem subject to a hard latency constraint per region and headroom ratio of replica utilization. This problem formulation is done through a mixed-integer linear programming (MILP) approach in which the units of the model are well specified and the feasibility verification process is just needed. Second, we design a simple greedy algorithm taking into account the assigned load and sorting the candidate clusters in descending order based on the increment of their cost. This cost is made up of the activation cost as well as the incremental cost due to replica installation per unit of available capacity along with the worst case of runtime. Third, we conduct extensive experiments, with 30 different seeds for each problem sizes, giving average, median, dispersion, and range of the optimality gap and runtime values.

\section{Background and Related Work}

\subsection{API Gateway Deployment and Governance}

A controlled benchmark of Kong Gateway on Amazon ECS versus EKS found that ECS Fargate offers faster cold starts and lower operational overhead, while EKS achieves higher peak throughput and more consistent latency under sustained load, at roughly 8\% higher monthly cost for the tested three-node configuration \cite{punniyamoorthy2025kong}. That work fixes the deployment to one cluster and compares orchestration platforms within it. Governance-aware, intent-driven architectures for multi-cluster API gateways address a different problem: once gateways are deployed across a fixed set of clusters, how security and performance policy are kept consistent and verifiable across them; the set of clusters is an input to that architecture, not a decision it makes. An SLO- and cost-aware autoscaling framework for Kubernetes adjusts pod replica counts within one cluster in response to live signals a runtime control problem distinct from the static, cross-cluster placement decision addressed here.

\subsection{Facility Location Foundations}

Capacitated facility location, deciding which candidate sites to open and how to assign demand to open sites to minimize total opening and delivery cost, is a classical NP-hard combinatorial problem. Distance- or latency-restricted variants appear in CDN replica placement,  cloud content placement and geographic $k$-center placement. Our formulation is a direct instance of this family, specialized to gateway replicas as the facility and request throughput as the capacity unit.

\subsection{Cloud and Edge Service Placement}

Volley formulates geo-distributed cloud data placement to jointly minimize latency and inter-datacenter cost \cite{agarwal2010volley}. Response-time-optimized distributed cloud resource allocation models a convex capacitated facility location problem with integrated queuing \cite{keller2016responsetime}; our model omits queuing delay (Section~VII) and targets a latency threshold rather than a response-time distribution. VNF placement in multi-cloud service chains solves a related facility-location-style problem \cite{bhamare2019vnf}. In the edge-cloud continuum, SLA-aware heuristic placement \cite{almeida2025tetris}, ILP-based QoS-aware placement \cite{hudson2021qosedge}, and learning-based latency-aware placement \cite{abedpour2026epnco} address closely related problems, situated within the broader fog/edge placement survey \cite{aitsalaht2020survey}. None targets API gateway replica placement specifically, and to our knowledge no prior formulation combines fixed activation cost, per-replica capacity, heterogeneous demand, and a strict per-region latency threshold in this domain.
\subsection{Kubernetes Multi-Cluster Scheduling}

Surveys of Kubernetes scheduling algorithms \cite{senjab2023kubescheduling} characterize generic, multi-objective, AI-based, and autoscaling-enabled scheduling strategies, but these operate at the level of assigning pods to nodes within already-provisioned clusters. The placement decision in this paper is upstream of that layer: it determines which clusters should exist in the gateway fleet at all, before any pod-to-node scheduling takes place within them.

\section{Problem Formulation}
\label{sec:problem}

Let $I$ be the set of client demand regions and $J$ the set of candidate clusters. For region $i \in I$: $d_i$ is request demand (RPS) and $\mathrm{SLA}_i$ is the maximum acceptable network round-trip latency (ms). For cluster $j \in J$: $f_j$ is fixed monthly activation cost (USD/month), $c_j$ is the monthly cost per gateway replica (USD/month), $\kappa_j$ is the sustainable request capacity of one replica (RPS/replica) at full utilization, and $r_j^{\max}$ is the maximum number of replicas permitted. A utilization headroom factor $0 < \rho \le 1$ (here $\rho = 0.8$) caps each replica's usable capacity at $\rho\kappa_j$, reserving margin for burst traffic and avoiding provisioning at benchmarked 100\% capacity. Let $\ell_{ij}$ (ms) be the estimated network latency between region $i$ and cluster $j$. The SLA-feasible set for region $i$ is $F_i = \{j \in J : \ell_{ij} \le \mathrm{SLA}_i\}$.

\emph{Feasibility.} A necessary condition for the instance to be feasible is that $F_i \neq \emptyset$ for every $i \in I$ and that aggregate SLA-restricted capacity is at least aggregate demand; this condition is not sufficient in general, since different regions' feasible sets can overlap and compete for the same restricted capacity, and final feasibility is certified by the solver's status, not asserted a priori. Every instance used in this paper's experiments is verified to satisfy $F_i \neq \emptyset$ for all $i$ before use (Section~V); no SLA relaxation or nearest-cluster fallback is applied anywhere in the model or in either solution method.

Decision variables: $x_j \in \{0,1\}$ (cluster $j$ activated), $y_j \in \mathbb{Z}_{\ge 0}$ (replicas in cluster $j$), and $z_{ij} \in [0,1]$ for $j \in F_i$ (fraction of region $i$'s demand served by cluster $j$; traffic may be split across multiple SLA-feasible clusters via weighted routing).

\begin{align}
\min \quad & \sum_{j \in J} \left( f_j x_j + c_j y_j \right) \label{eq:obj} \\
\text{s.t.} \quad & \sum_{j \in F_i} z_{ij} = 1 & \forall i \in I \label{eq:cover} \\
& \sum_{\substack{i \in I:\\ j \in F_i}} d_i\, z_{ij} \le \rho\, \kappa_j\, y_j & \forall j \in J \label{eq:cap} \\
& x_j \le y_j \le r_j^{\max} x_j & \forall j \in J \label{eq:link} \\
& x_j \in \{0,1\},\ y_j \in \mathbb{Z}_{\ge 0},\ z_{ij} \in [0,1] &
\end{align}

Constraint~\eqref{eq:cover} assigns every region's demand entirely to SLA-feasible clusters (infeasible pairs are excluded from the model by only defining $z_{ij}$ for $j \in F_i$, not by a penalty term). Constraint~\eqref{eq:cap} caps assigned load by usable replica capacity. Constraint~\eqref{eq:link} ties replica count to activation: an inactive cluster has zero replicas, an active one has at least one and at most $r_j^{\max}$. The resulting optimization is a mixed-integer capacitated facility-location problem. Its computational hardness follows from the underlying facility-location structure, for which NP-hard special cases are well established \cite{cornuejols1990uflp}.

\section{Proposed Approach}

\subsection{Exact MILP Solution}

The formulation above is solved with CBC 2.10.3 via PuLP 3.3.2, single-threaded, with a relative MIP-gap tolerance of 0.0 (i.e., the solver is required to certify true optimality, not merely a small gap) and a documented wall-clock time limit per instance (Section~V). We report the solver's returned status (\texttt{Optimal} or time-limit-reached) for every instance rather than assuming optimality, and we report the number of decision variables and constraints per instance size.

\subsection{Corrected Greedy Heuristic}

An earlier version of this heuristic computed available cluster capacity as $\kappa_j r_j^{\max} - \kappa_j y_j$, which ignored load already assigned to existing replicas' headroom, and ranked candidate clusters by per-replica cost alone, ignoring fixed activation cost. Both are corrected in Algorithm~\ref{alg:greedy}: available capacity is computed from assigned load $L_j$ directly against usable capacity $\rho\kappa_j r_j^{\max}$, replica count is recomputed by explicit ceiling division after each assignment, and candidates are ranked by total incremental cost (fixed cost, if not yet open, plus per-replica cost for any newly required replicas) divided by the load actually assignable to that candidate.

\begin{algorithm}[t]
\caption{Cost-Ordered Greedy Placement (corrected)}
\label{alg:greedy}
\begin{algorithmic}[1]
\Require Clusters $J$ with $f_j, c_j, \kappa_j, r_j^{\max}$; regions $I$ with $d_i$; feasible sets $F_i$; headroom $\rho$
\Ensure Opened clusters, replicas, assignment, cost, or \texttt{INFEASIBLE}
\State $L_j \gets 0,\ y_j \gets 0\ \forall j \in J$; sort $I$ by decreasing $d_i$
\For{each $i \in I$}
  \State $\mathit{rem} \gets d_i$
  \While{$\mathit{rem} > 0$ and a feasible candidate with spare capacity remains}
    \For{each $j \in F_i$ with $\rho\kappa_j r_j^{\max} - L_j > 0$}
      \State $\mathit{take} \gets \min(\mathit{rem},\ \rho\kappa_j r_j^{\max} - L_j)$
      \State $y_j' \gets \lceil (L_j + \mathit{take}) / (\rho\kappa_j) \rceil$; $\ \Delta y \gets y_j' - y_j$
      \State $\Delta C \gets f_j\mathbb{1}[x_j{=}0] + c_j \Delta y$; rank by $\Delta C / \mathit{take}$
    \EndFor
    \State Commit to the lowest-ratio candidate $j^\ast$: update $L_{j^\ast}$, $y_{j^\ast}$, open $j^\ast$ if needed, accumulate cost, $\mathit{rem} \mathrel{-}= \mathit{take}$
  \EndWhile
  \If{$\mathit{rem} > 0$} \Return \texttt{INFEASIBLE} (unassigned demand) \EndIf
\EndFor
\State \Return opened clusters, $y$, assignment fractions, total cost
\end{algorithmic}
\end{algorithm}

With $|I|$ regions and $|J|$ clusters, each region considers at most $|J|$ candidates per assignment round and requires at most $|J|$ rounds in the worst case, giving $O(|I||J|^2)$ worst-case time; in practice far fewer rounds are needed since most regions are satisfied by one or two clusters. The heuristic explicitly returns \texttt{INFEASIBLE} rather than silently leaving demand unassigned.

\section{Experimental Setup}

Candidate clusters are drawn from twelve real cloud provider regions across AWS, GCP, and Azure, each with a relative pricing multiplier (0.85-1.20) applied to a base fixed cost of \$72.00/month and a base per-replica cost of \$248.00/month, reflecting representative inter-region price variation, not dated list prices from any specific provider. Each replica's benchmarked capacity is $\kappa_j = 1200$ RPS, consistent with prior Kong Gateway benchmarking on comparable compute \cite{punniyamoorthy2025kong}, with headroom $\rho = 0.8$ and $r_j^{\max} = 20$. Demand regions correspond to ten geographic client population centers with $d_i \sim \mathrm{Uniform}\{200,\ldots,4000\}$ RPS and $\mathrm{SLA}_i$ drawn uniformly from $\{50, 75, 100, 150\}$ ms. Latency is synthesized as $\ell_{ij} = (5.0 + \mathrm{dist}_{ij}/100) \cdot U$, where $\mathrm{dist}_{ij}$ is the great-circle distance (km) between region and cluster coordinates and $U \sim \mathrm{Uniform}(0.9, 1.15)$ is multiplicative jitter; this is a stated modeling simplification (Section~VII), not measured production latency. All randomness (cluster/region draw, demand, jitter) is generated from a single seeded \texttt{random.Random(seed)} instance per problem instance, with no dependence on global random state, so a given seed reproduces an identical instance regardless of what else has executed. Every instance is verified strictly SLA-feasible before use; infeasible draws (rare) are discarded and the seed incremented, and the number of attempts required is logged.

One \emph{canonical instance} (10 clusters, 10 demand regions, requested seed 777, 3 attempts to reach a feasible draw) is used consistently for every baseline and sensitivity result in this paper; its full cluster and demand-region parameters are given in Table~\ref{tab:clusters}-\ref{tab:demands}, and its complete $10\times10$ latency matrix is provided in the accompanying data repository rather than reproduced in full here. Solver runs used CBC 2.10.3 via PuLP 3.3.2, single-threaded, on a single commodity cloud sandbox VM (x86\_64).

\begin{table}[t]
\caption{Canonical instance: candidate clusters}
\label{tab:clusters}
\centering
\scriptsize
\begin{tabular}{lrrrr}
\toprule
\textbf{Cluster} & \textbf{$f_j$ (\$)} & \textbf{$c_j$ (\$)} & \textbf{$\kappa_j$ (RPS)} & \textbf{$r_j^{\max}$} \\
\midrule
aws-us-east-1 & 72.00 & 248.00 & 1200 & 20 \\
aws-us-west-2 & 75.60 & 260.40 & 1200 & 20 \\
aws-eu-west-1 & 80.64 & 277.76 & 1200 & 20 \\
aws-ap-south-1 & 61.20 & 210.80 & 1200 & 20 \\
aws-ap-southeast-1 & 68.40 & 235.60 & 1200 & 20 \\
gcp-us-central1 & 70.56 & 243.04 & 1200 & 20 \\
gcp-europe-west1 & 79.20 & 272.80 & 1200 & 20 \\
gcp-asia-east1 & 64.80 & 223.20 & 1200 & 20 \\
azure-eastus & 73.44 & 252.96 & 1200 & 20 \\
azure-westeurope & 82.80 & 285.20 & 1200 & 20 \\
\bottomrule
\end{tabular}
\end{table}

\begin{table}[t]
\caption{Canonical instance: demand regions}
\label{tab:demands}
\centering
\scriptsize
\begin{tabular}{lrrr}
\toprule
\textbf{Demand region} & \textbf{$d_i$ (RPS)} & \textbf{SLA (ms)} & \textbf{Nearest cluster (ms)} \\
\midrule
north-america-east & 2506 & 50 & aws-us-east-1 (6.9) \\
north-america-west & 2435 & 100 & aws-us-west-2 (16.0) \\
europe-west & 3498 & 75 & gcp-europe-west1 (8.7) \\
europe-central & 3777 & 75 & gcp-europe-west1 (10.6) \\
asia-south & 914 & 100 & aws-ap-south-1 (8.9) \\
asia-east & 1849 & 50 & gcp-asia-east1 (13.4) \\
asia-southeast & 3732 & 50 & aws-ap-southeast-1 (8.7) \\
south-america & 2256 & 150 & azure-eastus (67.8) \\
oceania & 992 & 150 & aws-ap-southeast-1 (72.9) \\
middle-east & 1116 & 100 & aws-ap-south-1 (23.1) \\
\bottomrule
\end{tabular}
\end{table}

\section{Results}

\subsection{MILP vs. Greedy Heuristic (Repeated Trials)}

\begin{figure}[t]
\centering
\includegraphics[width=\linewidth]{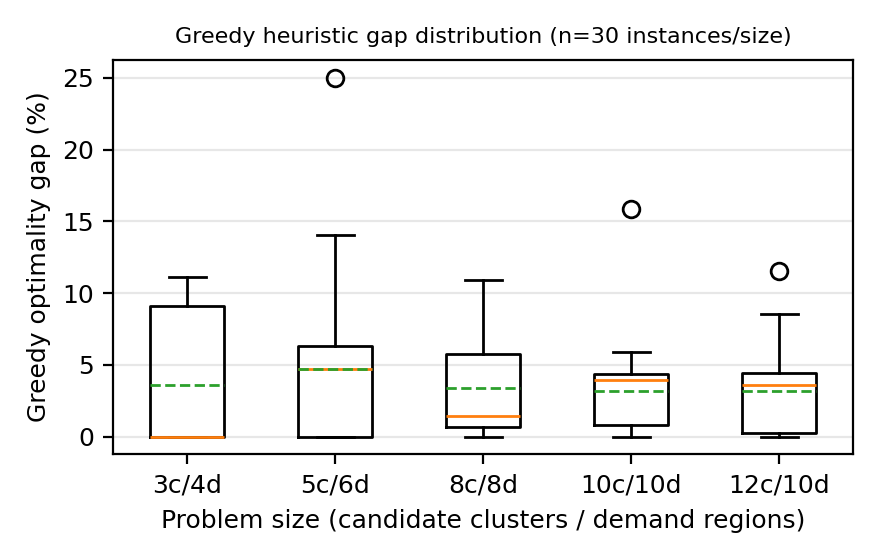}
\caption{Greedy optimality gap distribution across 30 independently seeded instances per problem size (box: quartiles; whiskers: 1.5$\times$IQR; solid line: median; dashed line: mean; circles: outliers).}
\label{fig:gap_dist}
\end{figure}

For five problem sizes, 30 independently seeded feasible instances were generated and solved by both methods. Table~\ref{tab:milp_greedy} reports aggregate statistics; CBC returned \texttt{Optimal} status for 100\% of all 150 instances. Fig.~\ref{fig:gap_dist} shows the full gap distribution per size as box plots rather than a single point estimate.

\begin{table}[t]
\caption{Greedy optimality gap and runtime vs. MILP (n=30 instances/size, CBC Optimal rate 100\% throughout)}
\label{tab:milp_greedy}
\centering
\scriptsize
\begin{tabular}{lrrrrr}
\toprule
\textbf{Size} & \textbf{Gap mean} & \textbf{Gap med.} & \textbf{Gap max} & \textbf{MILP $t$ med.} & \textbf{Ratio} \\
\midrule
3c/4d & 3.62\% & 0.00\% & 11.14\% & 9.7 ms & 662$\times$ \\
5c/6d & 4.73\% & 4.68\% & 25.02\% & 45.3 ms & 1772$\times$ \\
8c/8d & 3.41\% & 1.43\% & 10.92\% & 111.4 ms & 2155$\times$ \\
10c/10d & 3.20\% & 3.91\% & 15.88\% & 118.1 ms & 1437$\times$ \\
12c/10d & 3.20\% & 3.62\% & 11.51\% & 313.2 ms & 3489$\times$ \\
\bottomrule
\end{tabular}
\end{table}

The gap does not shrink monotonically with problem size (median ranges from 0.0\% at 3c/4d to 4.7\% at 5c/6d), and single-instance maxima up to 25\% occur even at small sizes, confirming the gap is driven by instance-specific cost/latency structure rather than scale alone. The measured MILP-to-greedy median runtime ratio ranges from approximately 660$\times$ to 3,490$\times$ across the tested sizes, roughly 2.8 to 3.5 orders of magnitude; we report this range rather than a single "three-to-four-orders-of-magnitude" figure, since the ratio itself varies by problem instance.

\subsection{Solver Scaling Behavior}

\begin{figure}[t]
\centering
\includegraphics[width=\linewidth]{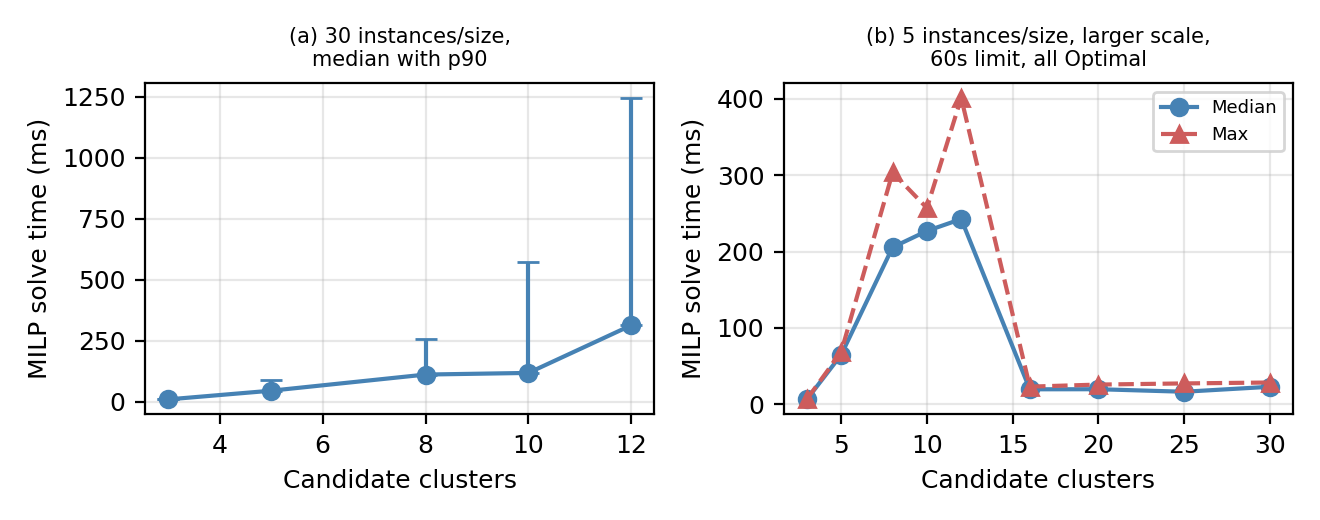}
\caption{(a) MILP median solve time with p90 error bars, 30 instances/size. (b) MILP solve time at larger candidate-cluster counts, 5 instances/size, 60-second limit; all instances solved to certified optimality.}
\label{fig:milp_time}
\end{figure}

Fig.~\ref{fig:milp_time}(a) shows MILP median solve time with p90 error bars for the sizes in Table~\ref{tab:milp_greedy}. Fig.~\ref{fig:milp_time}(b) extends this to larger candidate-cluster counts (up to 30, 5 instances/size, 60-second time limit), where CBC returned \texttt{Optimal} status for 100\% of instances at every size tested; no instance reached the time limit. Solve time is \emph{not} monotonic in cluster count (median peaks at 12 clusters, then drops sharply by 16), indicating that branch-and-bound difficulty in this problem family depends on instance-specific cost/latency/feasibility structure at least as much as raw size; we do not claim MILP solving becomes impractical at any tested scale, only that its cost grows with size on average.

The greedy heuristic was separately tested up to 80 candidate clusters (10 instances/size, Table~\ref{tab:greedy_scale}); its median runtime grows mildly and remains under 1.1~ms throughout, roughly linear in cluster count as expected from its $O(|I||J|^2)$ worst case with small realized constants.

\begin{table}[t]
\caption{Greedy heuristic runtime at larger scale (n=10 instances/size)}
\label{tab:greedy_scale}
\centering
\small
\begin{tabular}{rrr}
\toprule
\textbf{Clusters} & \textbf{Median time} & \textbf{Max time} \\
\midrule
12 & 0.082 ms & 0.098 ms \\
20 & 0.263 ms & 0.273 ms \\
30 & 0.390 ms & 0.400 ms \\
50 & 0.615 ms & 0.649 ms \\
80 & 1.026 ms & 1.034 ms \\
\bottomrule
\end{tabular}
\end{table}

\subsection{Baseline Comparison (Canonical Instance)}

\begin{figure}[t]
\centering
\includegraphics[width=0.85\linewidth]{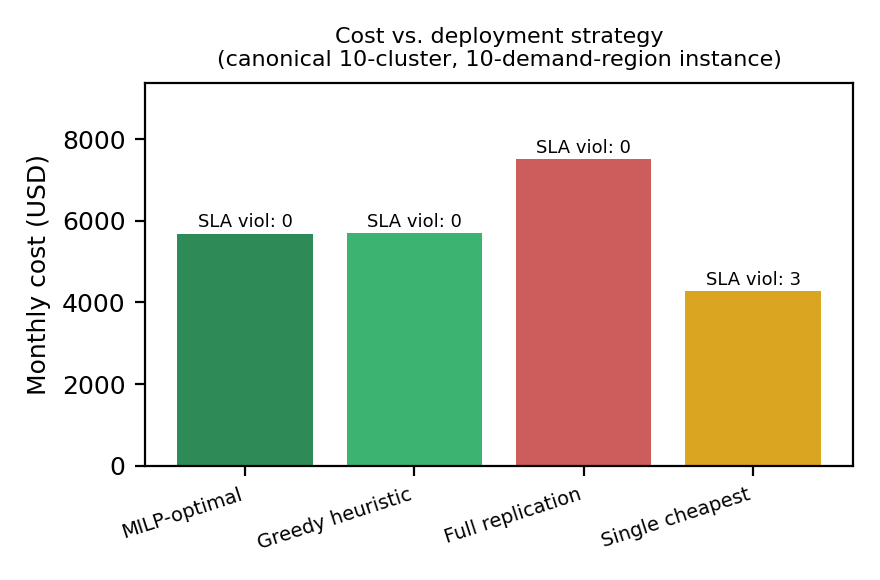}
\caption{Monthly cost and SLA-violation count across four deployment strategies on the canonical instance.}
\label{fig:baselines}
\end{figure}

We compare against two precisely defined baselines on the canonical instance. \emph{Full replication} opens every candidate cluster and assigns each demand region to its lowest-latency SLA-feasible open cluster (here, effectively its nearest feasible cluster, since all clusters are open); each cluster is then provisioned only for the load actually assigned to it, under the identical utilization headroom $\rho$ used by the proposed methods it does not duplicate global capacity at every cluster, which would not be a like-for-like comparison. \emph{Single cheapest} deploys only to the single lowest-per-replica-cost cluster regardless of latency. Table~\ref{tab:baselines} and Fig.~\ref{fig:baselines} report results; the MILP and greedy solutions activate the identical three clusters (\texttt{aws-ap-south-1}, \texttt{gcp-us-central1}, \texttt{gcp-asia-east1}), differing only in replica allocation.

\begin{table}[t]
\caption{Proposed methods vs. baselines (canonical instance)}
\label{tab:baselines}
\centering
\small
\begin{tabular}{lrrr}
\toprule
\textbf{Strategy} & \textbf{Cost/mo} & \textbf{Clusters} & \textbf{SLA viol.} \\
\midrule
MILP-optimal (proposed) & \$5{,}684.80 & 3 & 0 \\
Greedy (proposed) & \$5{,}697.20 & 3 & 0 \\
Full replication (baseline) & \$7{,}501.52 & 10 & 0 \\
Single cheapest (baseline) & \$4{,}277.20 & 1 & 3 \\
\bottomrule
\end{tabular}
\end{table}

On this instance, MILP-optimal placement costs 24.2\% less than full replication while both achieve zero SLA violations (full replication achieves this because every region's nearest feasible cluster is, by construction, among the opened set here; this need not hold in every instance). Single-cheapest is 24.8\% cheaper than the MILP-optimal placement but violates the latency SLA for 3 of 10 regions, illustrating that the cheapest deployment and the SLA-compliant deployment are not the same thing on this instance.

\subsection{SLA Sensitivity (Canonical Instance)}

\begin{figure}[t]
\centering
\includegraphics[width=0.85\linewidth]{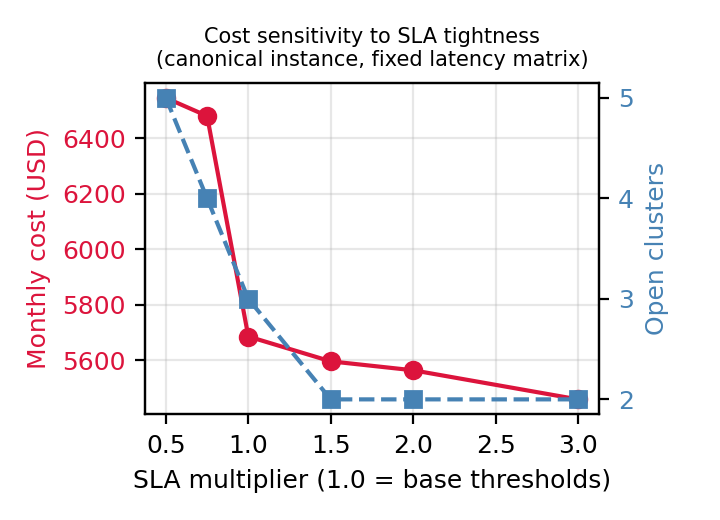}
\caption{Cost and open-cluster count as the SLA threshold is scaled on the canonical instance's fixed latency matrix.}
\label{fig:sla}
\end{figure}

The per-region SLA threshold was scaled by a multiplier from 0.5$\times$ to 3.0$\times$ on the \emph{same} canonical instance, i.e., the identical latency matrix, demand, and costs, varying only the SLA thresholds. Every tested multiplier remained feasible. Table~\ref{tab:sla} and Fig.~\ref{fig:sla} report the MILP-optimal cost and number of activated clusters at each multiplier.

\begin{table}[t]
\caption{SLA sensitivity, MILP-optimal (canonical instance, fixed latency matrix)}
\label{tab:sla}
\centering
\small
\begin{tabular}{rrrr}
\toprule
\textbf{SLA mult.} & \textbf{Cost/mo} & \textbf{Clusters} & \textbf{Status} \\
\midrule
0.5$\times$ & \$6{,}545.60 & 5 & Optimal \\
0.75$\times$ & \$6{,}480.48 & 4 & Optimal \\
1.0$\times$ (base) & \$5{,}684.80 & 3 & Optimal \\
1.5$\times$ & \$5{,}595.20 & 2 & Optimal \\
2.0$\times$ & \$5{,}562.96 & 2 & Optimal \\
3.0$\times$ & \$5{,}458.00 & 2 & Optimal \\
\bottomrule
\end{tabular}
\end{table}

On this instance, halving the SLA threshold (1.0$\times \to$ 0.5$\times$) raises cost by 15.1\% and increases activated clusters from 3 to 5; loosening it to 3$\times$ reduces cost by 4.0\% relative to the base threshold and settles at 2 clusters rather than collapsing to the single-cheapest baseline, since even a loose SLA does not make every region's nearest low-cost cluster mutually compatible with a single-cluster solution here. We report this as a finding specific to the evaluated instance, not a general property of all instances.

\section{Discussion}

For the evaluated instances, avoiding unnecessary cluster activation was the primary source of modeled cost reduction relative to full replication; fixed per-cluster activation cost, not marginal replica cost, was the larger contributor to full replication's cost premium. The greedy heuristic's 3-5\% mean, up to 25\% single-instance-maximum optimality gap is real and does not shrink predictably with scale, so its 660-3,490$\times$ runtime advantage should be weighed against solution quality per use case: for infrequent planning cycles, MILP's higher but still sub-second solve time is a reasonable cost for a certified-optimal answer; for frequent re-evaluation, the heuristic's speed may be worth its gap, particularly if periodically checked against an exact solve.

Translating a computed placement into a running deployment requires downstream automation outside this paper's scope: converting a placement into cluster-specific configuration and keeping it consistent with governance policy is the kind of problem governance-aware, intent-driven gateway architectures address, which could accept a computed placement as an input constraint. We do not model migration cost, rollout safety, or how often re-solving is warranted; the fast solve times measured here make frequent re-solving cheap, but detecting \emph{when} it is warranted is left to future work.

\section{Threats to Validity}
The latency matrix is synthesized from a great-circle propagation model with multiplicative jitter, not measured from live paths; it excludes routing asymmetry, congestion, and gateway/queueing/backend components (Section~\ref{sec:intro}). Demand, SLA targets, and cost multipliers are likewise synthetic, so costs should be read as normalized relative figures rather than list prices, and the 1200~RPS replica capacity derives from one prior benchmark under a specific configuration~\cite{punniyamoorthy2025kong}. Modeling demand as divisible via weighted traffic steering understates difficulty: a single-cluster-per-region constraint would require binary $z_{ij}$. The model also excludes egress cost, provider quotas, heterogeneous instance types, failures, redundancy constraints, and migration cost. The greedy gap is verified against an exact bound only up to 12 clusters, not at the scales in Table~\ref{tab:greedy_scale}. All figures are simulation outputs from one implementation on one hardware configuration.

\section{Conclusion and Future Work}

This paper formulated API gateway replica placement across multi-cloud Kubernetes clusters as a capacitated facility-location problem with a strict per-region latency threshold and replica utilization headroom, distinct from prior work on single-cluster gateway performance, post-deployment policy governance, and within-cluster autoscaling. An exact MILP formulation, solved to certified optimality with documented solver configuration, and a corrected greedy heuristic were evaluated across repeated, independently seeded synthetic instances. On the tested instances, the greedy heuristic's mean optimality gap was 3.2-4.7\% at a measured 660-3,490$\times$ runtime advantage; on one canonical instance used consistently throughout, MILP-optimal placement cost 24.2\% less than a precisely defined full-replication baseline with zero SLA violations, and tightening the latency SLA twofold raised cost by 15.1\% on that same instance. These results demonstrate the formulation's feasibility and the heuristic's practical trade-off but do not establish production cost or latency improvements. Future work includes validating the latency model against measured inter-region paths, replacing point demand estimates with time-varying traffic, extending the model to redundancy and failure-domain constraints, and integrating the placement decision as an input to governance-aware gateway configuration tooling.

\section*{Reproducibility Statement}
The instance generator, MILP and greedy implementations, canonical instance parameter files (Tables~\ref{tab:clusters}-\ref{tab:demands} plus the full latency matrix), solver metadata, and all raw per-trial CSV results underlying every table and figure are provided in the accompanying code repository bundled with this submission \cite{multicloud_kubernetes_code}.


\begin{thebibliography}{16}

\bibitem{kim2024kubeaegis}
B. Kim and S. Lee,
``KubeAegis: A Unified Security Policy Management Framework for
Containerized Environments,''
\emph{IEEE Access}, vol. 12, pp. 160636--160652, 2024,
doi: 10.1109/ACCESS.2024.3487990.

\bibitem{vu2022predictive}
D.-D. Vu, M.-N. Tran, and Y. Kim,
``Predictive Hybrid Autoscaling for Containerized Applications,''
\emph{IEEE Access}, vol. 10, pp. 109768--109778, 2022,
doi: 10.1109/ACCESS.2022.3214985.

\bibitem{marchese2025slo}
A. Marchese and O. Tomarchio,
``SLO-Aware Container Orchestration on Kubernetes Clusters,''
in \emph{Proc. IEEE 18th Int. Conf. Cloud Computing (CLOUD)},
2025, pp. 318--327,
doi: 10.1109/CLOUD67622.2025.00040.

\bibitem{cornuejols1990uflp}
G. Cornuejols, G. L. Nemhauser, and L. A. Wolsey, ``The uncapacitated facility location problem,'' in \emph{Discrete Location Theory}, P. B. Mirchandani and R. L. Francis, Eds. New York: Wiley, 1990, pp. 119-171.

\bibitem{sahoo2017replica}
J. Sahoo, M. A. Salahuddin, R. Glitho, H. Elbiaze, and W. Ajib, ``A survey on replica server placement algorithms for content delivery networks,'' \emph{IEEE Commun. Surveys Tuts.}, vol. 19, no. 2, pp. 1013-1044, 2017.

\bibitem{salahuddin2018contentplacement}
M. A. Salahuddin, J. Sahoo, R. Glitho, H. Elbiaze, and W. Ajib, ``A survey on content placement algorithms for cloud-based content delivery networks,'' \emph{IEEE Access}, vol. 6, pp. 91-114, 2018.

\bibitem{hillmann2020kcenter}
P. Hillmann, T. Uhlig, G. D. Rodosek and O. Rose, "A novel approach to solve K-center problems with geographical placement," 2015 IEEE International Conference on Service Operations And Logistics, And Informatics (SOLI), Yasmine Hammamet, Tunisia, 2015, pp. 31-36, doi: 10.1109/SOLI.2015.7367406.

\bibitem{punniyamoorthy2025kong}
V. Punniyamoorthy, S. R. Sankiti, N. Chockalingam, A. Agarwal, A. M. Kirubakaran, B. Kumar, K. Kannan, and S. Malempati, ``Analyzing performance and operational trade-offs in Kong Gateway deployments on AWS ECS and EKS platforms,'' in \emph{Proc. IEEE Int. Conf. Computer and Applications (ICCA)}, 2025.

\bibitem{agarwal2010volley}
S. Agarwal, J. Dunagan, and N. Jain, ``Volley: Automated data placement for geo-distributed cloud services,'' in \emph{Proc. 7th USENIX Symp. Networked Systems Design and Implementation (NSDI)}, 2010.

\bibitem{keller2016responsetime}
M. Keller and H. Karl, ``Response-time-optimized distributed cloud resource allocation,'' in \emph{Proc. ACM SIGCOMM Workshop on Distributed Cloud Computing (DCC)}, 2014, pp. 47-52; extended version \emph{arXiv preprint arXiv:1601.06262}, 2016.



\bibitem{bhamare2019vnf}
D. Bhamare, M. Samaka, A. Erbad, R. Jain, L. Gupta, and H. A. Chan, “Optimal virtual network function placement in multi-cloud service function chaining architecture,” Computer Communications, vol. 102, pp. 1–16, 2017, doi: 10.1016/j.comcom.2017.02.011.


\bibitem{almeida2025tetris}
L. Almeida and M. Peixoto, ``Tetris: An SLA-aware application placement strategy in the edge-cloud continuum,'' \emph{arXiv preprint arXiv:2511.00294}, 2025.

\bibitem{hudson2021qosedge}
N. Hudson, H. Khamfroush, and D. E. Lucani, ``QoS-aware placement of deep learning services on the edge with multiple service implementations,'' in \emph{Proc. IEEE ICCCN Workshop on Big Data and Machine Learning for Networking}, 2021.

\bibitem{abedpour2026epnco}
K. Abedpour, M. Garshasbi Herabad, Z. Li, and J. Taheri, ``EP-NCO: Latency-aware service placement using neural combinatorial optimisers for edge-cloud systems,'' \emph{arXiv preprint arXiv:2606.25553}, 2026.

\bibitem{aitsalaht2020survey}
F. Aït Salaht, F. Desprez, and A. Lebre, ``An overview of service placement problem in fog and edge computing,'' \emph{ACM Comput. Surv.}, vol. 53, no. 3, article 65, 2020.

\bibitem{senjab2023kubescheduling}
K. Senjab, S. Abbas, N. Ahmed, and A. U. R. Khan, ``A survey of Kubernetes scheduling algorithms,'' \emph{J. Cloud Comput.}, vol. 12, no. 1, article 87, 2023.

\bibitem{multicloud_kubernetes_code}
V. Punniyamoorthy, ``Multi-Cloud Kubernetes API Gateway Placement:
Source Code and Experimental Artifacts,'' 2026. [Online]. Available:
\url{https://github.com/Vinodhsrii/Multi-Cloud-Kubernetes}. Accessed: Aug. 26, 2026.

\end{thebibliography}
\end{document}